\documentclass[journal]{IEEEtran}
\usepackage{bm}
\usepackage{epsf}
\usepackage{times}
\usepackage{float}
\usepackage{graphicx}
\usepackage{amsmath,amssymb,amsthm}
\usepackage{color}
\usepackage{cite}
\usepackage{subcaption}
\usepackage{algorithm}
\usepackage[algo2e]{algorithm2e}
\usepackage[]{algpseudocode}

\makeatletter
\newcommand*{\rom}[1]{\expandafter\@slowromancap\romannumeral #1@}

\newtheorem{proposition}{Proposition}

\graphicspath{{subPert/}}

\def \bx{\bm x}
\def \bn{\bm n}
\def \bQ{\bm Q}
\def \bs{\bm s}

\def \bX{\bm X}
\def \by{\bm y}
\def \bz{\bm z}
\def \bv{\bm v}

\def \bQ{\bm Q}
\def \bB{\bm B}

\def \hatX{\hat{X}}

\DeclareMathOperator*{\tsum}{\textstyle\sum}

\begin{document}

\title{
{Communication efficient privacy-preserving distributed optimization using adaptive differential quantization}
\author{Qiongxiu Li,~\IEEEmembership{Student Member,~IEEE}, Richard Heusdens, ~\IEEEmembership{Senior Member,~IEEE},
and Mads~Gr\ae sb\o ll~Christensen,~\IEEEmembership{Senior Member,~IEEE}
\thanks{Q. Li and M. G. Christensen are with the Audio Analysis Lab, CREATE, Aalborg University, 9000 Aalborg, Denmark  (emails: \{qili,mgc\}@create.aau.dk).}
\thanks{R. Heusdens is with the Netherlands Defence Academy (NLDA), Het Nieuwe Diep 8, 1781 AC Den Helder, The Netherlands, and with the Faculty of Electrical Engineering, Mathematics and Computer Science, Delft University of Technology, Mekelweg 4, 2628 CD Delft, The Netherlands (email: r.heusdens@\{mindef.nl,tudelft.nl\}).}
}
}

\maketitle
\raggedbottom
\begin{abstract}
Privacy issues and communication cost are both major concerns in distributed optimization. There is often a trade-off between them because the encryption methods required  for privacy-preservation often incur expensive communication bandwidth. To address this issue, we, in this paper, propose a quantization-based approach to achieve both communication efficient and privacy-preserving solutions in the context of distributed optimization. By deploying an adaptive differential quantization scheme, we allow each node in the network to achieve its optimum solution with a low communication cost while keeping its private data unrevealed. Additionally, the proposed approach is general and can be applied in various distributed optimization methods, such as the primal-dual method of multipliers (PDMM) and the alternating direction method of multipliers (ADMM). Moveover, we consider two widely used adversary models: passive and eavesdropping.  Finally, we investigate the properties of the proposed approach using different applications and demonstrate its superior performance in terms of several parameters including accuracy, privacy, and communication cost.
\end{abstract}

\section{Introduction}
With the emergence of interconnected or networked systems, distributed optimization is widely used to process its massive amount of data. As the primary computation units in these distributed networks are often personal devices, such as mobile phones and tablets \cite{anderson2015technology,poushter2016smartphone}, the underlying networked data are private in nature. Furthermore, the available computational resources are also limited by the hardware and energy consumption. As a consequence, novel distributed optimization tools are required that are able to address the privacy concern in a way that is efficient in terms of communication and computational resources.

Existing approaches mostly address the above challenges only partially. To achieve computationally lightweight solutions, noise insertion approaches, which add noise to obfuscate the private data, are widely used in the literature. These methods can be broadly classified into three classes. The first one is the class of differentially private distributed optimization approaches \cite{huang2015differentially,han2016differentially,nozari2018differentially,zhang2016dynamic,zhang2018recycled,zhang2018improving,xiong2020privacy}. The main idea is to guarantee that the posterior guess of the private data is only slightly better than the prior guess. The downside of these algorithms is that they compromise the algorithm accuracy, i.e., they have an inherent trade-off between privacy and accuracy. Additionally, it is hard to achieve differential privacy in practice.  The second class is that of secret-sharing based distributed optimization approaches \cite{tjell2020privacy,tjell2019privacy} which deploy secret sharing to prevent privacy leakage, a technique used in secure multiparty computation \cite{Cramer2015,damgaard2012multiparty}. Secret sharing works by splitting the private data into a number of so-called shares and distributes them over the nodes such that without a sufficient number of nodes cooperating the private data cannot be reconstructed.  This, however, comes with additional communication costs as the distribution of shares requires extra communication. The third class is the class of subspace perturbation based distributed optimization approaches \cite{Jane2020ICASSP, Jane2020LS, Jane2020TSP} which, by inserting noise in a subspace determined by the graph topology, alleviates the privacy-accuracy trade-off without severely increasing the communication costs. 

Note that when considering the communication cost, aside from the number of times the communication channel is used, there is another critical parameter, namely the communication bandwidth or the corresponding bit-rate. The communication bandwidth is often omitted in privacy related approaches by assuming infinite precision. However, there is often a fundamental trade-off between privacy and communication cost in noise insertion type methods. The reason for this is that a higher privacy level usually requires a larger amount of noise insertion, which, in turn, increases the noise entropy and thereby the bit-rate. 
\subsection{Paper contributions}
The main contribution of this paper that we propose a new privacy-preserving distributed optimization approach to mitigate the aforementioned trade-off between privacy and communication bandwidth. To do so, we first give an information-theoretical analysis on this trade-off in noise insertion approaches. After that we propose to address this trade-off by exploiting an adaptive differential quantization scheme. To the best of our knowledge, this is the first approach which provides information theoretical privacy guarantees for distributed optimization with quantization. The proposed approach has a number of attractive properties:
\begin{itemize}
    \item It addresses the privacy and communication bandwidth trade-off by significantly reducing the overall communication cost compared to existing approaches including secret sharing, subspace perturbation and differential privacy based approaches. 
    \item The accuracy of the optimization result is not compromised by considering both privacy and quantization.
    \item It is generally applicable to various distibuted optimizers such as ADMM, PDMM and  related algorithms.
    \item It provides privacy guarantees under two widely-considered adversary models: eavesdropping and passive adversary models. 
\end{itemize}

\subsection{Outlines and notation}
The rest of the paper is organized as follows. Section \ref{section: preliminary} reviews fundamentals of distributed optimization. Section \ref{sec.proDef} introduces important concepts about privacy and then formulate the problem to be solved. Section \ref{sec.tradeoff} explains the reason why there is always a trade-off between privacy and communication bandwidth and Section \ref{sec.pro} introduces the proposed approach to address it. Numerical results and comparisons with existing approaches are demonstrated in Section \ref{sec.num}. Finally, the conclusion is given in Section \ref{sec.conclu}.

We use the following notation throughout this paper. Lowercase letters $x$ denote scalars, lowercase boldface letters $\bx$ denote vectors, uppercase boldface letters $\bX$ denote matrices.  $\bx_{i}$ and $\bm X_{ij}$ denote the $i$-th and  $(i,j)$-th entry of the vector $\bx$  and the matrix $\bX$, respectively. Denote calligraphic letters $\mathcal{X}$ as sets and  uppercase letters $X$ denote random variables having realizations $x$. $H(X)$ and $h(X)$ denote the Shannon entropy and differential entropy of a random variable $X$, respectively.


\section{Fundamentals}\label{section: preliminary}
This section reviews necessary fundamentals for distributed optimization.

\subsection{Distributed optimization over networks}\label{subsec:convex}
We model a distributed network as a graph $\mathcal{G}=(\mathcal{N},\mathcal{E})$  where  $\mathcal{N}={\{1,2,...,n}\}$ is the set of nodes and $\mathcal{E}\subseteq \mathcal{N}\times \mathcal{N}$ is the set of edges. 
Moreover, let $\mathcal{N}_{i}=\{j\,| \,{(i,j)\in \mathcal{E}\}}$ denote the set of neighboring nodes and $d_i=|\mathcal{N}_{i}|$ denote the degree (number of neighboring nodes) of node $i$. 
A standard distributed convex optimization problem with constraints over the network can  be formulated as
\begin{equation} 
\begin{array}{ll} &{\displaystyle \min_{\bx_i, \forall i\in \mathcal{N}}}  {{\displaystyle \sum_{i \in \mathcal{N}}} f_{i}(\bx_{i})} \\ 
&{\text {s.t.~}}  \forall (i,j)\in \mathcal{E} : {\bB_{i|j}\bx_{i}+\bB_{j|i}\bx_{j}=\bm b_{i,j}}
\end{array}\label{eq.setup}
\end{equation}
where $\bx_i\in \mathbb{R}^{u}$ denotes the optimization variable of node $i$, $ f_{i}: \mathbb{R}^{u}\mapsto \mathbb{R} \cup\{\infty\}$ denotes the local objective function which is assumed to be convex, $\bB_{i|j}$ and $\bB_{j|i}$ are edge-related matrices (weights) and $\bm b_{i,j}\in \mathbb{R}^{u}$ denotes the constraint imposed at edge $(i,j)\in \mathcal{E}$.  For simplicity, we assume $u=1$ and $\bm b_{i,j}=\bm 0$,  but the results can easily be generalized to arbitrary dimension and cases where $\bm b_{i,j}\neq\bm 0$. 
With this, $\bB_{i|j}$ and $\bB_{j|i}$ are related to entries of the incidence matrix $\bB\in \mathbb{R}^{m\times n}, m=|\mathcal{E}|,$ of the graph: $\bB_{li}=\bB_{i|j}=1$, $\bB_{lj}=\bB_{j|i}=-1$  if and only if $e_l = (i,j)\in \mathcal{E}$ and $i<j$,  
\subsection{Distributed optimizers}
\label{sec:distr}
To solve the problem in \eqref{eq.setup} in a decentralized manner, i.e., where each node is only allowed to exchange information with its neighboring nodes, a number of distributed, iterative optimizers have been proposed, including ADMM \cite{boyd2011distributed} and PDMM \cite{biADMM2015,zhang2018distributed, sherson2018derivation}. 
It has been shown using  monotone operator theory and operator splitting techniques that ADMM and PDMM are closely related \cite{sherson2018derivation} (see \cite{ryu2016primer} for details on monotone operator theory). The updating equations at iteration $t=0,1,\hdots$ can be generally represented as
\begin{align*}
&\bx_{i}^{(t+1)} = \arg\min_{\bx_{i}} \big(
 f_{i}(\bx_{i})+ \tsum_{j \in \mathcal{N}_i} \bz_{i|j}^{(t)} \bB_{i|j}\bx_i + \frac{cd_i}{2}\bx_{i}^2 \big) 
 \\
    &\forall j \in \mathcal{N}_{i}: \bz_{j|i}^{(t+1)}=\theta\bz_{j|i}^{(t)}+(1-\theta) ({\bz}_{i|j}^{(t)}+2c\bB_{j|i}\bx_i^{(t+1)}), 
\end{align*}
where $c$ is a constant for controlling the convergence rate. Each edge $e_l=(i,j)\in\mathcal{E}$ is associated to two auxiliary variables $\bz_{}=\bz_{i|j}$ and $\bz_{l+m}=\bz_{j|i}$. Stacking all auxiliary variables together we have $\bz\in \mathbb{R}^{2m}$. 
$\theta\in [0,1)$ is a constant for controlling the averaging of the nonexpansive operators. For example, $\theta=0$ results in Peaceman-Rachford splitting (PDMM) and $\theta=1/2$ results in Douglas-Rachford splitting (ADMM).
For simplicity,  we will use  $\theta=0$, i.e., PDMM, as an example to explain the main idea but the conclusions hold for all $\theta\in [0,1)$.

\section{Problem definition}\label{sec.proDef}
In this section we first introduce important concepts regarding privacy-preservation and then define the problem to be solved and its evaluation metrics. 
\subsection{Privacy concern}
In distributed optimization, sensitive personal information is often embedded in each node's local objective function $f_{i}(\bx_i)$. The main reason is that the local objective function contains node-specific data as input and such data are often private in nature.  As an example, consider a smart grid application. Assume each household/node in the network has its own power consumption data $\bs_i$ and the goal is to compute the global average of these power consumption data, i.e., $n^{-1}\sum_{i \in \mathcal{N}} \bs_i$. In this context the local objective function is given by $f_{i}(\bx_i)=\frac{1}{2}\|\bx_i - \bs_i\|^2_2$ and the overall problem setup can be formulated as follows:
\begin{equation} \label{eq.setupAve}
\begin{array}{ll}{\displaystyle \min_{\bx_i}} & {{\displaystyle\sum_{i \in \mathcal{N}} \frac{1}{2}\|\bx_i - \bs_i\|^2_2}} \\ {\text { s.t. }} & {\bx_i=\bx_j, {\forall} (i,j)\in \mathcal{E}}.\end{array}
\end{equation} 
Note that the power consumption data $\bs_i$, contained in the local objective function $f_{i}(\bx_i)$,  of each household should be protected from being revealed to others, because it can reveal information regarding the householders such as their activities and even their health conditions like whether they are disabled or not \cite{giaconi2018privacy}. 
Hence,  information regarding the local objective function $f_{i}(\bx_i)$ is considered to be sensitive and should be  protected from being revealed in the process of solving the optimization problem. 

\subsection{Adversary model}
\begin{figure}[t]
\centering
\includegraphics[width=.45\textwidth]{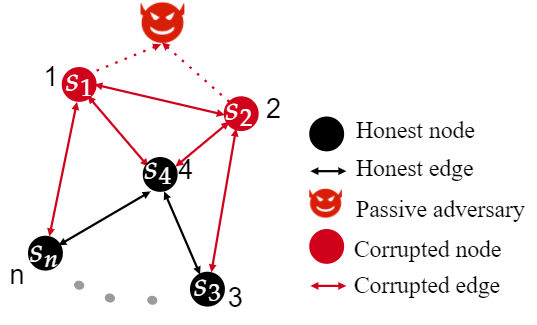}
 \centering
\caption{Passive adversary}
\label{fig.pasAdv}
\vskip -10pt
\end{figure}

To analyze the privacy, we must specify an adversary model. The purpose of such a model is to quantify the system robustness in dealing with different security attacks.  In this paper, we consider two types of adversary models: the passive and the eavesdropping model.  The passive (also called semi-honest or honest-but-curious) adversary model is a classical model considered in distributed networks \cite{bogdanov2008sharemind}. It works by colluding a number of nodes, referred to as corrupted nodes. The corrupted nodes are assumed to follow the algorithm instructions (called the protocol) but will share information together.  We call an edge in the graph corrupted as long as  there is one corrupted node at its ends. All messages transmitted along such an edge will be known to the corrupted nodes, thus also to the passive adversary. See Fig.\ref{fig.pasAdv} for a toy example.  Hence, the passive adversary will collect all information from the corrupted nodes to infer private data of the other non-colluding nodes (referred to as honest nodes).

The eavesdropping adversary is assumed to attack the system by listening to the messages transmitted along the communication channels, i.e., edges. This model receives little attention as it can be addressed by assuming all communication channels are securely encrypted such that  the transmitted  messages cannot be eavesdropped, e.g., secret sharing based approaches \cite{li2019privacyS,tjell2019privacy,tjell2020privacy}. However,  this  assumption  is particularly  expensive  to  realize  in  distributed  optimization applications, as a large number of iterations is often required. 

We assume that the considered two adversaries can cooperate, i.e., they share information together with the aim of inferring the private data of the honest nodes. 

\subsection{Main requirements and related metrics}\label{subsec.req}
Putting things together, we now state the main requirements  that communication efficient privacy-preserving distributed optimization should 
satisfy and introduce the related metrics. 
\begin{enumerate}
    \item \textbf{Output correctness}:  Each node $i$ should obtain the optimal solution to \eqref{eq.setup}, denoted by $\bx_i^{*}$, at the end of the algorithm. A typical way to quantify the  output correctness is to adopt certain distance metrics to calculate the difference between the estimated output $\bx^{(t)}$ and  the optimum output $\bx^{*}$. In this paper we use the overall mean square error (MSE) to quantify it, i.e., $\left\|\bx^{(t)}-\bx^{*}\right\|^{2}$.
    \item \textbf{Communication cost}: After the algorithm execution, the cost of all communications should be as low as possible. The communication cost is given by $T(2m)l$, where $T$ is the total number of iterations, $2m$ is the total amount of messages transmitted at each iteration ($d_i$ per node) and $l$ is the number of bits needed to represent each message.
    \item \textbf{Individual privacy}: Each node's private information, embedded in $f_{i}(\bx_i)$, should be protected under both eavesdropping and passive adversaries throughout the algorithm. As we are focusing on noise insertion approaches, we will focus on  information-theoretical metrics to describe the privacy.  In the context of distributed processing, two commonly used metrics are the so-called $\epsilon$-differential privacy and mutual information \cite{cover2012elements}.  In this paper we choose mutual information as the individual privacy metric. The main reasons are as follows.  (1) Mutual information has been proven effective in the context of privacy-preserving distributed processing \cite{Jane2020TIFS}, and has been applied in various applications \cite{lopuhaa2019information,Jane2020GSP,yagli2020information,bu2020tightening,pensia2018generalization,negrea2019information}. (2) It is closely related to $\epsilon$-differential privacy (see \cite{cuff2016differential} for more details), and is easier to realize in practice \cite{gotz2011publishing,haeberlen2011differential,korolova2009releasing}.  
    (3) $\epsilon$-differential privacy does not work if the private data is correlated \cite{kifer2011no}. 
\end{enumerate}

\section{Trade-off in noise insertion approaches}\label{sec.tradeoff}
In this section, we aim to show that there is typically a trade-off between privacy and communication bandwidth in noise insertion approaches. To do so we will first explain a simple noise insertion scheme and then introduce how to compute the communication bandwidth, i.e., bit-rate, after applying quantization. Finally, we give an example to demonstrate this trade-off.

\subsection{Additive noise insertion}
Assume a scenario that a number of people, each having his/her own private data,  would like to participate in a project which takes the private data of all these participants as input. Let $s$ denote the private data held by you and you are reluctant to share your private data to others due to privacy concerns.  The idea of noise insertion is to insert certain noise, denoted by $r$, to  obfuscate the private data and then share the obfuscated data, denoted by $s_r$, to others.  One of the most simple yet widely-used way of noise insertion is to directly add the noise to the private data for protecting it from being revealed to others, referred to as additive noise insertion, and defined as 
\begin{align}\label{eq.addiNoi}
    s_r=s+r.
\end{align}
Intuitively, a higher privacy level will be achieved if the obfuscated data $s_r$ is less correlated with the private data $s$. We have the following result. 
\begin{proposition}(Privacy guarantee for additive noise insertion)\label{prop.1} 
Let $R$ and $S$ be continuous random variables with variance $\sigma^{2}_{R},\sigma^{2}_{S}< \infty$, denoting the inserted noise and private data, respectively, and assume that $R$ and $S$ are statistically independent. Let $S_r=S+R$. Given an arbitrarily small $\delta>0$, there exists 
$\beta > 0$ such that for $\sigma^{2}_{R} \geq \beta$
\begin{align} \label{eq.delta}
    I(S;S_r)\leq \delta.
\end{align}
In the case that the noise $R$ is Gaussian distributed, we have
\begin{align}\label{eq.sigmaR}
    \beta =  \frac{\sigma^{2}_{S}}{2^{2\delta}-1}.
\end{align}
\begin{proof}
See \cite[Proposition $1$]{Jane2020TIFS}. 
\end{proof}
\end{proposition}
Hence, the more noise is inserted, the higher privacy level can be obtained. However, we remark that more noise will inevitably increase the noise entropy and thus requires a higher bit-rate (i.e., communication bandwidth). In what follows we will explain this in more detail.

\subsection{Quantization and bit-rate}
The main idea of quantization is to establish a mapping of the input data to a countable set of reproduction values, which is referred to as a codebook. More specifically, a quantizer divides the input domain into so-called quantization cells (Voronoi regions) where all elements within a cell are represented by a unique reproduction or code value. Let $l$ denote the number of bits to represent the reproduction values, $2^l$ in total.  Although there exists many different quantization schemes, we will introduce a simple yet effective one, namely  uniform quantization.  In this quantizer, all quantization cells have the same size, except for the cells at the boundary of the domain in the case of fixed bit-rate quantization. For example, a one-dimensional 2-bit uniform mid-rise quantizer with cell-width $\Delta$ will divide the input range into four regions, each represented by a unique code value. That is, the quantization cells are given by $(-\infty,-\Delta], (-\Delta,0], (0, \Delta],(\Delta,\infty)$ and respresented by $-\frac{3\Delta}{2},-\frac{\Delta}{2},\frac{\Delta}{2}$, and $\frac{3\Delta}{2}$, respectively.

Using a finite number of bits to quantize a continuous random variable  will always produce an error, or distortion. Intuitively, the fewer bits are used, the more distortion will occur. To have an idea on how many bits are required for transmitting a message, we can determine its entropy since this gives a lower bound on the number of bits needed to represent the data.  With a uniform quantizer, the entropy of a  quantized continuous random variable $X$ at sufficiently high rate can be approximated as \cite{gaard2007Quan}:
\begin{align}\label{disEn}
    H(\hatX) \approx h(X)-\frac{1}{2} \log \left(12 D_{\Delta_x}\right),
\end{align}
where $\hatX$ is the discrete random variable after quantizing $X$,  $\Delta_x$ denotes the quantization cell width and  $D_{\Delta_x}=\frac{\Delta_x^2}{12}$ is the quantization noise variance. Here, $H(\hatX)$ is the Shannon entropy of $\hatX$, and $h(X)$ is the differential entropy of $X$, assuming it exists. 
Since the differential entropy of a random variable with fixed variance $\sigma^2$ is upper bounded by 
\begin{align}
         h(X)\leq \frac{1}{2} \log \left(2 \pi e \sigma^{2}\right),   
\end{align}
we have 
\begin{align}\label{eq.disEnUpper}
    H(\hatX) \leq \frac{1}{2} \log \left(\frac{2 \pi e \sigma_{X}^{2}}{\Delta_x^2}\right).
\end{align}

\subsection{Trade-off between privacy and bits}
\begin{figure}
  \centering
  \includegraphics[width=.40\textwidth]{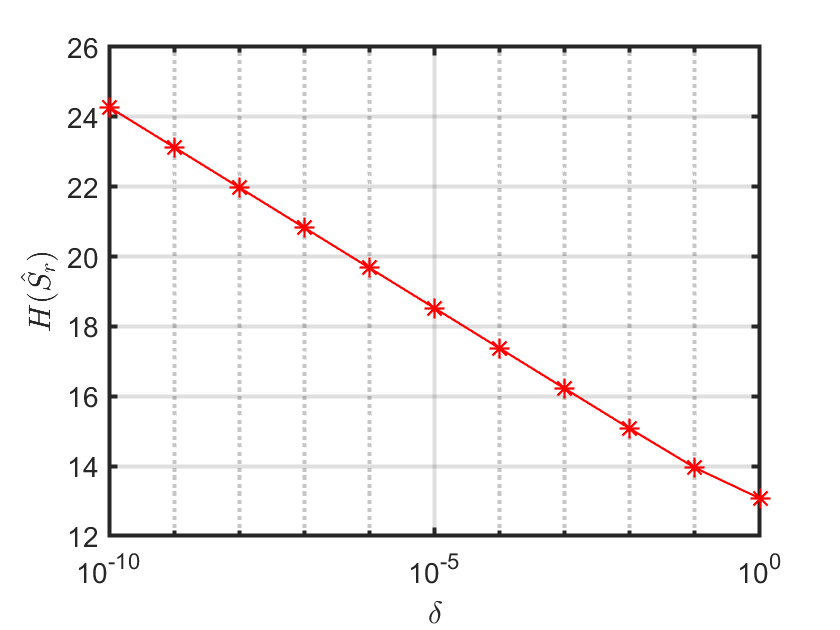}
  \caption{Bits for transmitting the obfuscated data $S_r$ in terms of the privacy level $\delta$.}
  \label{fig.entHsr}
\end{figure}

Assume the inserted noise $R$ is Gaussian distributed and the desired privacy level $\delta$  is given. The entropy of the obfuscated data $S_r$ is then given by 
\begin{align}
H(\hat{S_r})&\leq\frac{1}{2} \log \left(\frac{2 \pi e \sigma_{S_r}^{2}}{\Delta_{s_r}^2}\right) \nonumber \\
& \stackrel{{\text{(a)}}}{=}  \frac{1}{2} \log \left(\frac{2 \pi e (\sigma^{2}_{S}+\sigma^{2}_{R})}{\Delta_{s_r}^2}\right) \nonumber\\
&\stackrel{{\text{(b)}}}{=}  \frac{1}{2} \log \left(\frac{2 \pi e (2^{2\delta}\sigma^{2}_{S})}{(2^{2\delta}-1)\Delta_{s_r}^2}\right) \label{eq.bitHsr},
\end{align}
where (a) holds as $S$ and $R$ are independent and (b) follows by setting $\sigma^{2}_{R}$ equal to $\beta$ given by \eqref{eq.sigmaR}.
By inspection of \eqref{eq.bitHsr}, we can see that the smaller the information leakage $\delta$ is, i.e., the higher the privacy level is, the higher the amount of bits for representing the quantized obfuscated data $\hat{S_r}$ will be. Clearly there is a trade-off between them. In Fig. \ref{fig.entHsr} we give an example based on \eqref{eq.bitHsr} to demonstrate this trade-off, where we set $\sigma^{2}_{S}=1$ and  $\Delta_{s_r}=10^{-5}$.


\section{Proposed approach}\label{sec.pro}
After explaining why there is a trade-off between privacy and communication cost in noise insertion approaches, we now proceed to introduce the proposed approach which addresses this trade-off by using the adaptive differential quantization scheme of \cite{schellekens2017quantisation,jonkman2018quantisation}. The reason of choosing this quantization scheme is because it can help to save the communication cost without compromising privacy. In addition, we exploit the idea of additive noise insertion to guarantee privacy after applying quantization.  

In this section we will first briefly introduce the adaptive differential quantization technique. After that, we introduce the proposed approach and explain how to guarantee privacy by making use of the auxiliary variable $\bz^{(0)}$, see Section~\ref{sec:distr}, as noise. Finally, we analyze the performance of the proposed approach based on the concerned requirements mentioned in Section \ref{subsec.req}, i.e.,  individual privacy, output correctness and communication cost.

The main idea of applying adaptive differential quantization is based on the observation that
for fixed point iterations the difference of successive iterations will converge to zero (i.e, it is a Cauchy sequence), which implies that the entropy of the difference of successive iterations will decrease to zero as the number of iteration increases. Motivated by this, the adaptive differential quantization scheme proposed in \cite{schellekens2017quantisation,jonkman2018quantisation} quantize the difference of the auxiliary variable at every two successive iterations with an adaptive cell-width decreasing with increasing iterations, details will be given below. By doing so, low data rate transmission between nodes can be achieved without compromising the accuracy of the algorithm. 

With the adaptive differential quantization scheme,  the process of the proposed approach is given as follows. At initialization $t=0$, each node $i\in \mathcal{N}$ initializes its auxiliary variables $\{\bz_{j|i}^{(0)}\}_{j \in \mathcal{N}_{i}}$ and sends the corresponding $\bz_{j|i}^{(0)}$ to each and every neighboring node $j \in \mathcal{N}_{i}$, respectively. Then update $\bx_{i}^{(1)}$ and $\bz_{j|i}^{(1)}$ as 
\begin{align}
&\bx_{i}^{(1)} = \arg\min_{\bx_{i}} \big(
 f_{i}(\bx_{i})+ \tsum_{j \in \mathcal{N}_i} {\bz_{i|j}^{(0)}}\bB_{i|j}\bx_i + \frac{cd_i}{2}\bx_{i}^2 \big) \label{eq.xup0} \\
    &\forall j \in \mathcal{N}_{i}: \bz_{j|i}^{(1)}=\bz_{i|j}^{(0)}+2c\bB_{j|i}\bx_i^{(1)}  \label{eq.zup0} 
\end{align}
Let $\hat{\bz}$ denote the quantized version of $\bz$. At iteration $t\geq1$, each node does not transmit the unquantized $\bz_{j|i}^{(t)}$ to node $j$ directly, instead it first defines the difference variable $\bv^{(t)}$ as
\begin{align} \label{eq.vdef}
\bv^{(t)} \triangleq \left\{\begin{array}{ll}
\bz^{(1)} - \bz^{(0)}, & \textrm{ if $t=1$},\\
\bz^{(t)} - \hat{\bz}^{(t-1)}, & \textrm{ if $t>1$}.
\end{array}\right.
\end{align}
Let $Q(\cdot)$ denote the quantization operation. Applying quantization to the difference variable $\bv^{(t)}$ we have
\begin{align}\label{eq.vnoise}
    \hat{\bv}^{(t)} = Q(\bv^{(t)})
    = \bv^{(t)} +\bn_{q, v^{(t)}},
\end{align}
where $\bn_{q, v^{(t)}}$  denotes the noise introduced by quantizing $\bv^{(t)}$. The adaptive differential quantizater quantizes the difference variable  $\{\bv^{(t)}\}_{t\geq 1}$ with a geometrically decreasing cell-width $\Delta^{(t)}=\gamma^{t-1}\Delta^{(0)}$, where $\Delta^{(0)}$ denotes the initial cell-width and $\gamma\in (0,1)$ denotes the rate of growth.
After obtaining $\hat{\bv}^{(t+1)}$, the quantized $\hat{\bz}^{(t+1)}$ can be obtained by
\begin{align} \label{eq.zRec}
\hat{\bz}^{(t)}=\left\{\begin{array}{ll}
\bz^{(0)}+\hat{\bv}^{(1)}, & \textrm{ if $t=1$},\\
\hat{\bz}^{(t-1)} + \hat{\bv}^{(t)}, & \textrm{ if $t>1$}.
\end{array}\right.
\end{align}
Note that all $\{\hat{\bz}^{(t)}\}_{t\geq1}$ can be reconstructed when knowing $\bz^{(0)}$ and the quantized $\{\hat{\bv}^{(t)}\}_{t\geq 1}$ as 
\begin{align}
    \forall t\geq 1: \hat{\bz}^{(t)}=\bz^{(0)} + \sum_{\tau=1}^{t}\hat{\bv}^{(\tau)}\label{eq.vztau}.
\end{align}
After constructing $\hat{\bz}^{(t)}$, each node 
can  update the local variables $\bx_{i}^{(t+1)}$ and  $\bz_{j|i}^{(t+1)}$ using the quantized $\hat{\bz}_{i|j}^{(t)}$ from the previous iteration, i.e.,
\begin{align}
&\bx_{i}^{(t+1)} = \arg\min_{\bx_{i}} \big(
 f_{i}(\bx_{i})+ \tsum_{j \in \mathcal{N}_i} {\hat{\bz}_{i|j}^{(t)}}\bB_{i|j}\bx_i + \frac{cd_i}{2}\bx_{i}^2 \big) \label{eq.xup} \\
    &\forall j \in \mathcal{N}_{i}: \bz_{j|i}^{(t+1)}=\hat{\bz}_{i|j}^{(t)}+2c\bB_{j|i}\bx_i^{(t+1)}  \label{eq.zup} 
\end{align}

Overall, we conclude that all messages that need to be transmitted are  the initialized $\bz^{(0)}$ and the quantized $\{\hat{\bv}^{(t)}\}_{t\geq 1}$. Because all $\{\hat{\bz}^{(t)}\}_{t\geq 1}$ can be computed using $\bz^{(0)}$ and $\{\hat{\bv}^{(t)}\}_{t\geq 1}$, all $\{\bx^{(t)}\}_{t\geq 1}$ can be computed using  $\bz^{(0)}$ and $\{\hat{\bz}^{(t)}\}_{t\geq 1}$ using \eqref{eq.xup0} and \eqref{eq.xup}.

Having introduced how to apply adaptive differential quantization to reduce the communication bandwidth, we now proceed to explain how to guarantee privacy. Motivated based on the idea of using additive noise insertion to achieve privacy-preservation, instead of inserting extra noise we propose to make use of the initialized $\bz^{(0)}$ as noise. To guarantee privacy, we assume that it  is transmitted at high precision (see the following individual privacy analysis for details).  


 \begin{algorithm}[t]
    \caption{Communication efficient privacy-preserving distributed optimization using adaptive differential quantization}
    \label{alg:pro}
    \SetKwInOut{Input}{Input}
    \SetKwInOut{Output}{Output}
    \SetKwInOut{Initialization}{Initialization}
    \Input{Initialized auxiliary variables: $\{\bz_{j|i}^{(0)}\}_{(i,j) \in \mathcal{E}}$\\ Initial quantization cell-width: $\Delta^{(0)}$\\ 
    Number of bits of quantizer: $l$\\
    Rate of growth: $\gamma$ 
    } 
    \Output{Optimum optimization solutions: $\{\bx_{i}^{*}\}_{i\in\mathcal{N}}$}
      \Initialization{Each node $i\in \mathcal{N}$, initializes $\{\bz_{j|i}^{(0)}\}_{j \in \mathcal{N}_{i}}$ based on the desired privacy level.}
            \If {$\|\bx_i^{(t)} - \bx_i^*\|^2 < $ threshold}
            {
            \If{$t=0$} 
            { Receive $\bz_{i|j}^{(0)}$ from all neighbors $j\in \mathcal{N}_{i}$ through securely encrypted channels \cite{dolev1993perfectly}.\\
            Update $\bx_{i}^{(1)}$ using \eqref{eq.xup0},  $\{\bz_{j|i}^{(1)}\}_{j \in \mathcal{N}_{i}}$  using \eqref{eq.zup0}. \
            }
            \If{$t\geq 1$}
            { Receive $\hat{\bv}_{i|j}^{(t)}$ from all neighbors $j \in \mathcal{N}_{i}$ through non-securely encrypted channels. \\
            Update $\{\hat{\bz}_{i|j}^{(t)}\}_{j \in \mathcal{N}_{i}} $ using $ \eqref{eq.zRec}$. \\ \
             Update $\bx_{i}^{(t+1)}$ using  \eqref{eq.xup}, $\{\bz_{j|i}^{(t+1)}\}_{j \in \mathcal{N}_{i}}$ using \eqref{eq.zup}. \
             }
             Compute $\{\bv_{j|i}^{(t+1)}\}_{j \in \mathcal{N}_{i}}$ using  \eqref{eq.vdef}.\\ \
               Quantize $\{\bv_{j|i}^{(t+1)}\}_{j \in \mathcal{N}_{i}} $ with a $l$-bit uniform quantizer having cell-width $\Delta^{(t+1)}=\gamma^{t}\Delta^{(0)}$ 
              .\\ \
             Transmit $\hat{\bv}_{j|i}^{(t+1)}$ to each neighbor $j \in \mathcal{N}_{i}$.\
            }
\end{algorithm}


\begin{figure*}[ht]
\begin{subfigure}{0.50\textwidth}
\includegraphics[width=0.95\linewidth]{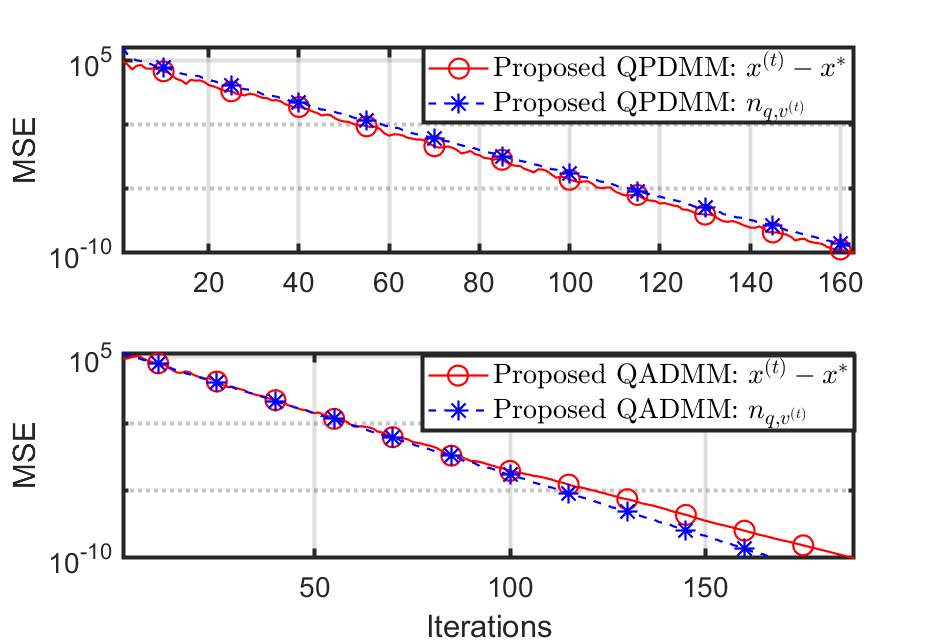} 
\caption{Distributed average consensus}
\end{subfigure}
\begin{subfigure}{0.50\textwidth}
\includegraphics[width=0.95\linewidth]{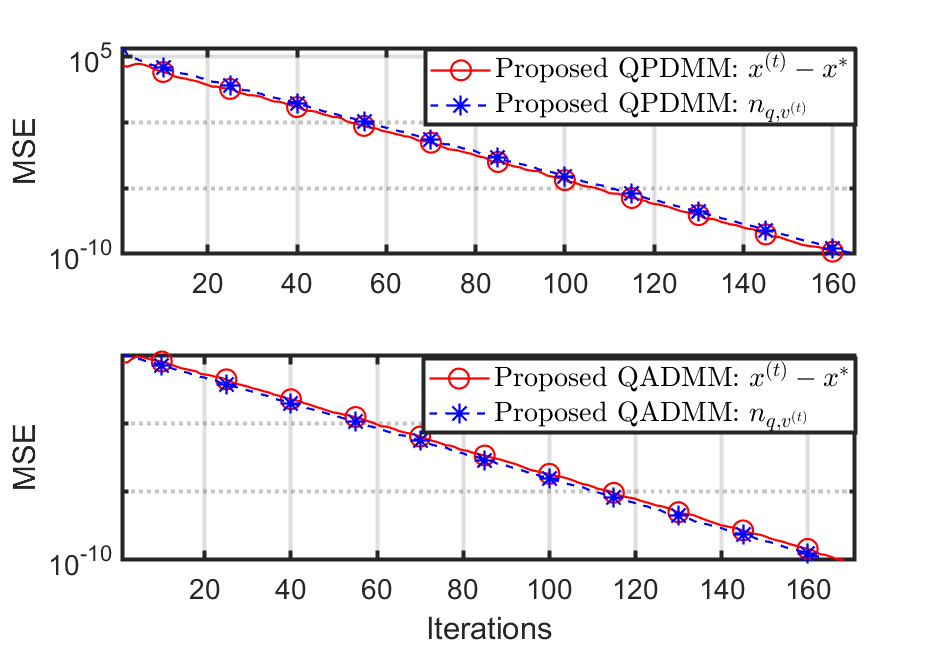}
\caption{Distributed least squares}
\end{subfigure}
\caption{Output correctness (MSE) of both $\bx^{(t)}-\bx^{*}$ and $\bn_{q, v^{(t)}}$ in terms of iteration numbers using the proposed quantized PDMM and ADMM algorithm (QPDMM and QADMM, respectively) for the distributed average consensus  and distributed least-squares applications.}
\label{fig.proAccu}
\end{figure*}

\subsection{Individual privacy}
By inspection of \eqref{eq.xup}, for $t\geq1$ the updates  $\bx_i^{(t)}$  satisfy
\begin{align}
   \partial f_{i}(\bx_i^{(t+1)})+\sum_{j \in \mathcal{N}_i} \bB_{i|j}\hat{\bz}_{i|j}^{(t)} +cd_i\bx_i^{(t+1)}=0. \label{eq.xgraDMM}
    \end{align}
 We can see that the private data is only contained in the subgradient $\partial f_{i}(\bx_i^{(t+1)})$. As a consequence, the goal of the privacy analysis is to see what information regarding  $\partial f_{i}(\bx_i^{(t+1)})$ is revealed during the iterations. Note that for $t=0$ we have $\partial f_{i}(\bx_i^{(1)})+\sum_{j \in \mathcal{N}_i} \bB_{i|j}\bz_{i|j}^{(0)} +cd_i\bx_i^{(1)}=0 $. 

For simplicity, assume $\bB_{i|j}=1$ for all $j \in \mathcal{N}_i$, and thus $\bB_{j|i}=-1$.
Denote $\mathcal{N}_c$ and $\mathcal{N}_h$ as the set of corrupted nodes and honest nodes, respectively. 
Let $\mathcal{N}_{i,c}=\mathcal{N}_i \cap \mathcal{N}_c$ and $\mathcal{N}_{i,h}=\mathcal{N}_i \cap \mathcal{N}_h$ denote the set of the corrupted and honest neighbors of the node $i$, respectively.  In addition, we assume a worse case that each honest node has at least one corrupted neighboring node, i.e., $\mathcal{N}_{i,c} \neq\emptyset$. Combining \eqref{eq.vdef} and \eqref{eq.vnoise} we conclude that $\hat{\bv}^{(t)}-\bn_{q, v^{(t)}}=\bz^{(t)}-\hat{\bz}^{(t-1)}$, so that \eqref{eq.zRec} can be expressed as   
\begin{align}
   \hat{\bz}^{(t)} &= \bz^{(t)} +\bn_{q, v^{(t)}}. \label{eq.nonqzv}
\end{align}
For node $k\in \mathcal{N}_{i,c}$, using \eqref{eq.zup} and \eqref{eq.nonqzv}, we can express the left-hand side of \eqref{eq.xgraDMM} as  
\begin{align}\label{eq.fxnCon}
    \partial f_{i}(\bx_i^{(t+1)})\!+\!\sum_{j\in \mathcal{N}_{i}} \hat{\bz}^{(t)}_{i|j}- \frac{d_i (\hat{\bz}_{k|i}^{(t+1)}\!-\bn_{q, v_{k|i}^{(t+1)}}\!-\hat{\bz}_{i|k}^{(t)}}{2}).
\end{align}  
To quantify the amount of information  about the private data $\partial f_{i}(\bx_i^{(t+1)})$ learned by the adversaries, we must first inspect what information is available to them.   We first consider the passive adversary.  As it can collect all the information available to the corrupted nodes, it has the following knowledge:
\begin{align}
 \{\bx_i^{(t)}\}_{i\in \mathcal{N}_c, t\geq 1}\cup \{{\bz}_{i|j}^{(0)}, \hat{\bv}_{i|j}^{(t+1)} \}_{(i,j) \in \mathcal{E}_c, t\geq 0}\nonumber,
\end{align}
where $\mathcal{E}_c=\{(i,j) \in \mathcal{E}, (i,j)\notin \mathcal{N}_h\times \mathcal{N}_h\}$ denotes the set of corrupted edges. 
With the above knowledge, the passive adversary is able to compute both $\sum_{j\in \mathcal{N}_{i,c}}\hat{\bz}^{(t)}_{i|j}$ using \eqref{eq.vztau} and $\frac{1}{2}d_i ( \hat{\bz}_{k|i}^{(t+1)}-\hat{\bz}_{i|k}^{(t)} )$ in \eqref{eq.fxnCon}. After computing these known terms, the unknown terms in \eqref{eq.fxnCon} are given by
\begin{align}\label{eq.passive}
 \{ &\partial f_{i}(\bx_i^{(t+1)})+\sum_{j\in \mathcal{N}_{i, h}} \hat{\bz}^{(t)}_{i|j}+\frac{d_i }{2} \bn_{q, v_{k|i}^{(t+1)}}\}_{k\in \mathcal{N}_{i,c}} .
\end{align} 
Next, we consider the eavesdropping adversary. 
To minimize the expense of requiring securely encrypted communication channels,  we propose to use secure channel encryption only once. More specifically, no channel encryption is involved except for transmitting  $\bz^{(0)}$ during the initialization step.
As a consequence, the eavesdropping adversary can listen to all transmitted messages after initialization, i.e., 
\begin{align}
  \{\hat{\bv}_{i|j}^{(t+1)} \}_{(i,j) \in \mathcal{E},t\geq0} \nonumber,
\end{align}
note that it does not have knowledge about ${\bz}_{i|j}^{(0)}$.
Based on  \eqref{eq.vztau}, we can, therefore, deduce $\sum_{\tau=1}^{t}\hat{\bv}_{i|j}^{(\tau)}$ from  $\hat{\bz}^{(t)}_{i|j}$ in \eqref{eq.passive} as it is known to the eavesdropping adversary. Consequently, we conclude that all what the passive and eavesdropping adversaries observe about the honest node $i$ is given by 
\begin{align} \label{eq.pFinal}
  \{\partial f_{i}(\bx_i^{(t+1)})+\sum_{j\in \mathcal{N}_{i, h}} {\bz}_{i|j}^{(0)}-\frac{d_i}{2}\bn_{q, v_{k|i}^{(t+1)}}\}_{k\in \mathcal{N}_{i,c}} 
\end{align}
where the last term $\{\bn_{q, v_{k|i}^{(t+1)}}\}_{j\in \mathcal{N}_{i,c}}$ will converge to the all-zero vector as the iterations proceed. 
If node $i$ has at least one honest neighbor, i.e., $\mathcal{N}_{i,h}\neq \emptyset$, the term $\sum_{j\in \mathcal{N}_{i, h}} \bz_{i|j}^{(0)}$ can be considered as noise. Hence, 
we can, by Proposition \ref{prop.1}, protect the private data $\partial f_{i}(\bx_i^{(t+1)})$ from being revealed by making the variance of $\bz^{(0)}$ arbitrarily large at the initialization step. Therefore, arbitrarily small information leakage regarding $\partial f_{i}(\bx_i^{(t+1)})$ can be achieved at every iteration.

We remark that the proposed quantized approach can be seen as a privacy-enhanced version of the existing subspace perturbation based approach \cite{Jane2020TSP} since it introduces an  additional   noise component $\bn_{q, v_{k|i}^{(t+1)}}$ in  \eqref{eq.pFinal} which helps to further protect the private data. 
Hence, we conclude that the conditions to guarantee the privacy of the data of honest node $i\in \mathcal{N}_h$  for both  passive and eavesdropping  adversaries are given by:
\begin{itemize}
    \item It has at least one honest neighbor. That is, $\mathcal{N}_{i,h}\neq \emptyset$.
    \item The communication channels are securely encrypted in the initialization phase when transmitting $\bz^{(0)}$. 
\end{itemize}

\begin{figure*}[ht]
\begin{subfigure}{0.50\textwidth}
\includegraphics[width=0.9\linewidth]{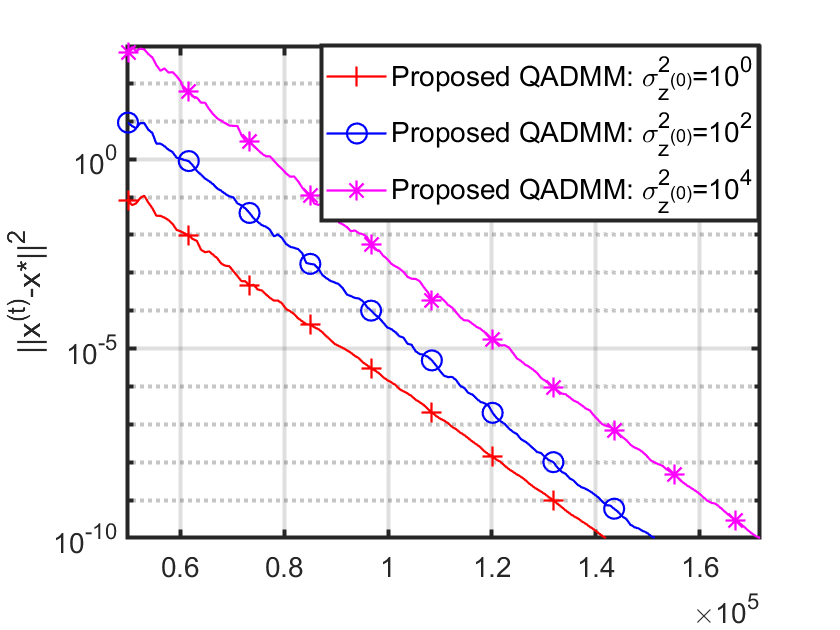} 
\caption{Proposed QADMM}
\end{subfigure}
\begin{subfigure}{0.50\textwidth}
\includegraphics[width=0.9\linewidth]{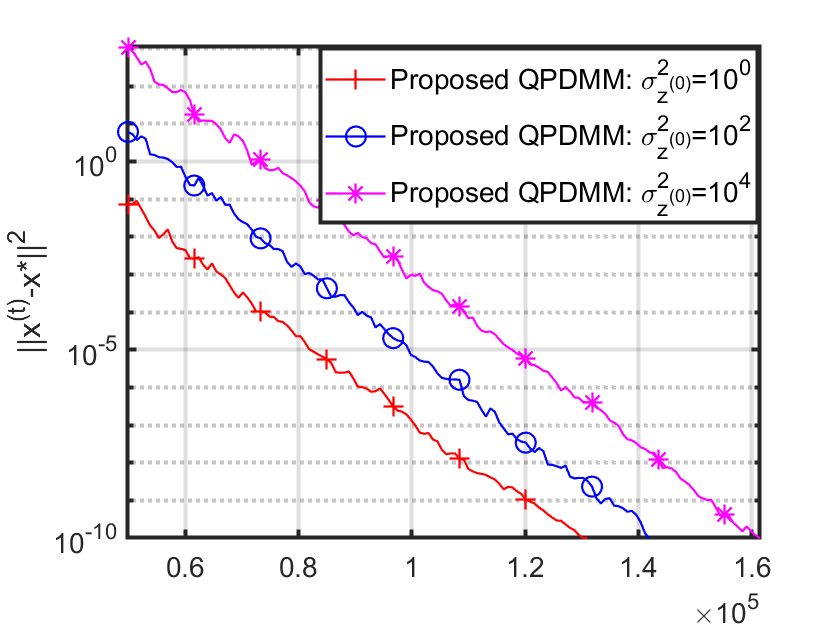}
\caption{Proposed QPDMM}
\end{subfigure}
\caption{MSE in terms of communication cost (the amount of bits) of the proposed QADMM and QPDMM for three different noise levels $\sigma^2_{z^{(0)}}$ of the  auxiliary variables in distributed average consensus application.}
\label{fig.proBit}
\end{figure*}

\begin{figure}
  \centering
  \includegraphics[width=.48\textwidth]{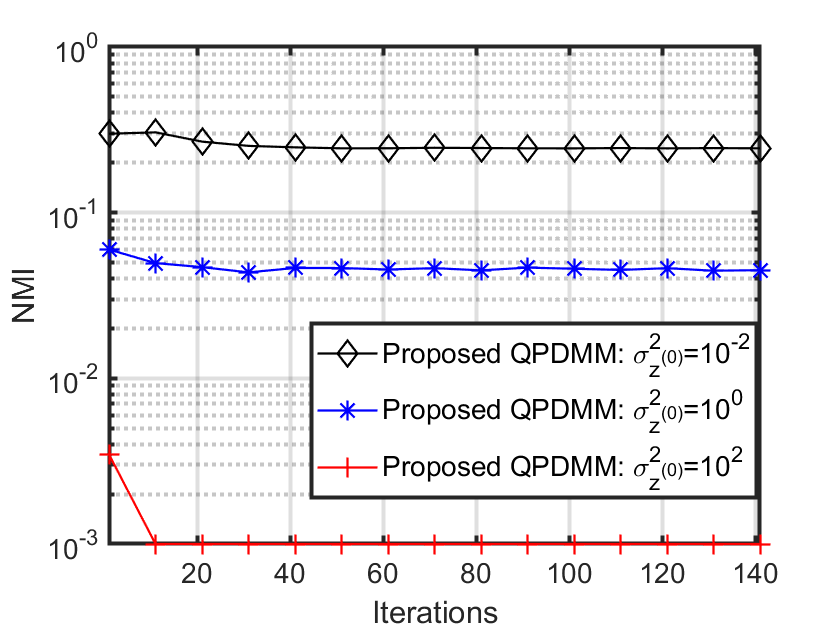}
  \caption{Individual privacy (normalized mutual information (NMI)) of the proposed QPDMM in terms of iteration numbers for  three different noise levels in distributed average consensus application.}
  \label{fig.proPri}
\end{figure}


\subsection{Output correctness and communication cost}\label{subsec.utility}
In \cite{jonkman2018quantisation} it has been shown that if the sequence $\{{\bn_{q, v^{(t)}}}\}_{t >0}$ is finitely summable, then Douglas-Rachford splitting will convergence to a fixed point $\bx^{*}$ which is the solution to  \eqref{eq.setup}. Hence, the output correctness requirement is satisfied.  As for the communication cost, we can see that with the help of quantization, the communication cost of the proposed approach is substantially reduced. In fact, as we will demonstrate in the next section, we can quantize the data with a one-bit quanitzer ($l=1$) and still obtain perfect output correctness and individual privacy.

 The details of the proposed algorithm are summarized in Algorithm \ref{alg:pro}.

\section{Numerical results}\label{sec.num}
In this section, we will demonstrate simulation results for the proposed approach and compare this with existing approaches.  We simulated a geometric network with $n=30$ nodes where every two nodes are allowed to transmit messages if their distance is within a radius of $\sqrt{\frac{2\log (n)}{n}}$, as this condition ensures that the corresponding graph is connected with high probability \cite{dall2002random}. 


\subsection{Performance of the proposed approach}
We use two applications to test the performance of the proposed approach: distributed average consensus and distributed least squares. The main reason for choosing these two applications is that they are intensively investigated in the literature \cite{gupta2017privacy,gupta2019statistical,li2019privacyA,li2019privacyS,tjell2019privacy,nozari2017differentially,he2019privacy,ruan2017secure,mo2017privacy,Jane2020ICASSP,Jane2020LS}. The detailed problem formulation of distributed average consensus is already introduced in \eqref{eq.setupAve}. As for distributed least squares, assume each node has partial knowledge of a linear system (assuming overdetermined) including an input observation, denoted as $\bQ_i \in \mathbb{R}^{p_{i} \times u}, \, p_i>u$ and a decision vector, denoted as $\by_i\in \mathbb{R}^{p_{i}}$.  Stacking the partial knowledge together we denote $\bQ=[\bQ^{\top}_1,\dots,\bQ^{\top}_n]^{\top} \in \mathbb{R}^{P_n \times u}$ and $\by=[\by^{\top}_1,\dots,\by^{\top}_n]^{\top}\in \mathbb{R}^{P_n}$, where $P_n=\sum_{i\in \mathcal{N}} p_{i}$. The goal of privacy-preserving distributed least squares is to allow each node to achieve the global optimum solution $\forall i\in \mathcal{N}, \,\bx_i^{*}=(\bQ^{\top}\bQ)^{-1}\bQ^{\top}\by \in \mathbb{R}^{u}$, without revealing its private data, i.e., $\bQ_i, \by_i$. With distributed optimization this problem can be formulated as 
\begin{equation} \label{eq.setupLS}
\begin{array}{ll}{\displaystyle \min_{\bx_i}} & {\sum_{i \in \mathcal{N}} \frac{1}{2}\|\by_{i}-\bQ_i\bx_{i}\|^{2}_{2}} \\ {\text { s.t. }} & {\bx_i=\bx_j, {\forall} (i,j)\in \mathcal{E}}.\end{array}
\end{equation}

 In all experiments, we randomly draw the private data, i.e., $\bs_i$ in the case of distributed average consensus and $\bQ_i, \by_i$ in the case of distributed least squares, from a zero-mean Gaussian distribution with unit variance. In addition, we set $c=\gamma=0.9$ and the auxiliary variable $\bz^{(0)}$ is initialized with zero-mean Gaussian distributed noise having a variance $\sigma^2_{z^{(0)}}\!={\Delta^{(0)}}^{2}$, where $\Delta^{(0)}$ is the initial quantization cell-width. Moreover, for the proposed quantized approach,  a one-bit (mid-rise) quantizer is used with cell-width $\Delta^{(t)}$, which means that we only transmit the signs of the $\bz_{i|j}$s which will be reconstructed at the receiver by $\pm \Delta^{(t)}/2$.

The performance of the proposed approach is demonstrated in terms of the three requirements mentioned in Section \ref{subsec.req}.
\begin{enumerate}
    \item \textbf{Output correctness}: From Fig.\ref{fig.proAccu}   
we can see that applying the proposed adaptive differential quantization scheme to both PDMM (QPDMM) and ADMM (QADMM), the quantization noise $\bn_{q, v^{(t)}}$ converges to zero and the optimization variable  $\bx^{(t)}$ converges to the optimal $\bx^{*}$. These results validate the claim stated in Section \ref{subsec.utility}: if the sequence $\{{\bn_{q, v^{(t)}}}\}_{t >0}$ is finite summable, the output correctness will be guaranteed.  Hence, we conclude that the proposed approach satisfies the output correctness requirement, i.e., accuracy is not compromised by considering both quantization and privacy. Additionally, it is generally applicable to both ADMM and PDMM.  

\item \textbf{Communication cost}: Fig. \ref{fig.proBit} demonstrates the total communication cost (the amount of bits) of the proposed QADMM and QPDMM under three different privacy levels: $\sigma^2_{z^{(0)}}=10^0, \sigma^2_{z^{(0)}}=10^2, \sigma^2_{z^{(0)}}=10^4$.  Note that for the proposed approach we have $l=1$ since a one-bit quantizer is used. We can see that the convergence rate of the proposed approach is invariant to the privacy level. 

\item \textbf{Individual privacy}: Fig.\ref{fig.proPri} shows the individual privacy of an arbitrary honest node  over iterations using the proposed approach under the condition that there is only one honest neighboring node when applied to the distributed average consensus problem. That is, the normalized mutual information measured based on \eqref{eq.pFinal} when $\mathcal{N}_{i,c}=d_i-1$. We can see that the larger $\sigma^2_{z^{(0)}}$ is, the less individual privacy is revealed, i.e., the higher the privacy level is.  Hence, the proposed approach is able to guarantee individual privacy by controlling the variance of the initialized $\bz^{(0)}$, i.e., $\sigma^2_{z^{(0)}}$. 
\end{enumerate}

\begin{figure}
  \centering
  \includegraphics[width=.48\textwidth]{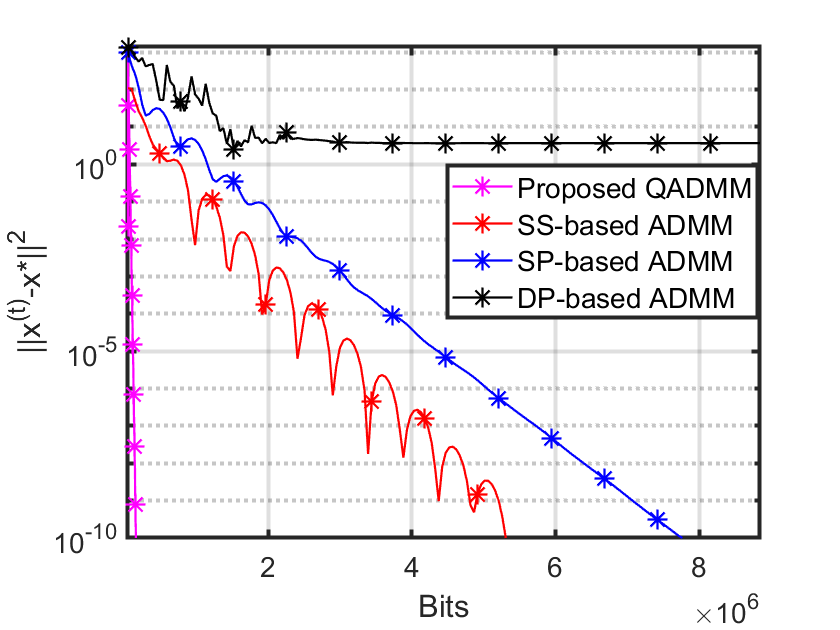}
  \caption{Communication cost (bits) comparisons of the proposed QADMM algorithm and the existing subspace perturbation approach \cite{Jane2020TSP}, secret sharing approach \cite{gupta2019statistical} and differential privacy approach \cite{nozari2017differentially} under the same privacy level  in distributed average consensus application.}
  \label{fig.comBit}
\end{figure}

\subsection{Comparison with existing approaches}
We now compare the performance of the proposed QADMM with existing approaches including subspace perturbation (SP) based approach \cite{Jane2020TSP}, secret sharing (SS) based approach \cite{gupta2019statistical} and differential privacy (DP) based approach \cite{nozari2017differentially}. To ensure a fair comparison, we insert the same amount of noise in all these algorithms. More specifically, all inserted noise are Gaussian distributed with zero mean and variance $10^2$. Additionally, all algorithms are based on the ADMM optimizer. 
Note that since none of the existing algorithms use a quantization scheme, the number of bits to represent each massage is set to $l=64$  (MATLAB double precision floating-point format).  From Fig.\ref{fig.comBit},  we can see that 
\begin{itemize}
    \item among these approaches, differential privacy based approach does not output an accurate result, i.e., it suffers from a privacy-accuracy trade-off;
    \item as expected, compared to all existing approaches the proposed algorithm significantly reduces the communication cost.
\end{itemize}
One remark is placed here. Among all existing approaches: subspace perturbation, secret sharing and differential privacy based approaches, the proposed approach is obtained by applying adaptive differential quantization to the subspace perturbation based approaches such that the communication cost is reduced without sacrificing the individual privacy and output correctness. For the other two types of approaches it would also be interesting to investigate how quantization affects their performances in terms of privacy and accuracy. 




\section{Conclusion}\label{sec.conclu}
 In this paper, we proposed a novel yet general communication efficient privacy-preserving distributed optimization approach using adaptive differential quantization.  By adopting an adaptive quantizer  that dynamically decreases its cell-width for each iteration to reduce the communication cost and making use of additive noise insertion to achieve privacy-preservation, we are able to alleviate the trade-off between privacy and communication cost without compromising the algorithm accuracy.  In addition,  the proposed algorithm is able to protect privacy of any honest node against the passive adversary by requiring only one honest neighboring node. Moreover, the proposed method is computationally very lightweight in its way of dealing with an eavesdropping adversary as no secure encryption is needed, except for in the initialization step. Finally, numerical results were conducted, which confirm the desirable properties of the proposed approach in terms of accuracy, privacy and communication cost and show that the proposed approach has superior performance compared to the existing approaches.  

\newpage
\bibliographystyle{IEEEbib}
\bibliography{dualpath}

\begin{thebibliography}{10}

\bibitem{anderson2015technology}
{M. Anderson},
\newblock {\em Technology device ownership, 2015},
\newblock Pew Research Center, 2015.

\bibitem{poushter2016smartphone}
{J. Poushter and others},
\newblock ``Smartphone ownership and internet usage continues to climb in
  emerging economies,''
\newblock {\em Pew Research Center}, vol. 22, pp. 1--44, 2016.

\bibitem{huang2015differentially}
{Z. Huang, S. Mitra, and N. Vaidya},
\newblock ``Differentially private distributed optimization., pp. 1--10,''
\newblock in {\em Proc. Int. Conf. Distrib. Comput. Netw}, 2015.

\bibitem{han2016differentially}
{S. Han, U. Topcu, and G. J. Pappas},
\newblock ``Differentially private distributed constrained optimization,''
\newblock {\em IEEE Trans. Autom. Control., vol. 62, no. 1, pp 50-64}, 2016.

\bibitem{nozari2018differentially}
{E. Nozari, P. Tallapragada, and J. Cort{\'e}s},
\newblock ``Differentially private distributed convex optimization via
  functional perturbation,''
\newblock {\em IEEE Trans. Control Netw. Syst., vol. 5, no. 1, pp 395-408},
  2018.

\bibitem{zhang2016dynamic}
{T. Zhang and Q. Zhu},
\newblock ``Dynamic differential privacy for {ADMM}-based distributed
  classification learning,''
\newblock {\em IEEE Trans. Inf. Forensics Security, vol. 12, no. 1, pp.
  172--187}, 2016.

\bibitem{zhang2018recycled}
{X. Zhang, M. M. Khalili, and M. Liu},
\newblock ``Recycled {ADMM}: Improve privacy and accuracy with less computation
  in distributed algorithms,''
\newblock in {\em in Proc. 56th Annu. Allerton Conf. Commun., Control, Comput.
  pp.959--965}, 2018.

\bibitem{zhang2018improving}
{X. Zhang, M. M. Khalili, and M. Liu},
\newblock ``Improving the privacy and accuracy of {ADMM}-based distributed
  algorithms,''
\newblock {\em Proc. Int. Conf. Mach. Lear. pp. 5796--5805}, 2018.

\bibitem{xiong2020privacy}
{Y. Xiong, J. Xu, K. You, J. Liu and L. Wu},
\newblock ``Privacy preserving distributed online optimization over unbalanced
  digraphs via subgradient rescaling,''
\newblock {\em IEEE Trans. Control Netw. Syst.}, 2020.

\bibitem{tjell2020privacy}
{K. Tjell and R. Wisniewski},
\newblock ``Privacy preservation in distributed optimization via dual
  decomposition and {ADMM},''
\newblock in {\em Proc. IEEE 58th Conf. Decis. Control., pp. 7203--7208}, 2020.

\bibitem{tjell2019privacy}
{K. Tjell, I. Cascudo and R. Wisniewski},
\newblock ``Privacy preserving recursive least squares solutions,''
\newblock in {\em Proc. Eur. Control Conf., pp.3490--3495}, 2019.

\bibitem{Cramer2015}
{R. Cramer, I. B. Damg{\aa}rd, and J. B. Nielsen},
\newblock {\em Secure Multiparty Computation and Secret Sharing},
\newblock Cambridge University Press, 2015.

\bibitem{damgaard2012multiparty}
{I. Damg{\aa}rd, V. Pastro, N. Smart, and S. Zakarias},
\newblock ``Multiparty computation from somewhat homomorphic encryption,''
\newblock in {\em Advances in Cryptology--CRYPTO, pp. 643--662}. Springer,
  2012.

\bibitem{Jane2020ICASSP}
{Q. Li, R. Heusdens and M. G. Christensen},
\newblock ``Convex optimisation-based privacy-preserving distributed average
  consensus in wireless sensor networks,''
\newblock in {\em Proc. Int. Conf. Acoust., Speech, Signal Process., pp.
  5895-5899}, 2020.

\bibitem{Jane2020LS}
{Q. Li, R. Heusdens and M. G. Christensen},
\newblock ``Convex optimization-based privacy-preserving distributed least
  squares via subspace perturbation,''
\newblock in {\em Proc. Eur. Signal Process. Conf., pp. 2110-2114}, 2021.

\bibitem{Jane2020TSP}
{Q. Li, R. Heusdens and M. G. Christensen},
\newblock ``Privacy-preserving distributed optimization via subspace
  perturbation: A general framework,''
\newblock in {\em IEEE Trans. Signal Process., vol. 68, pp. 5983 - 5996}, 2020.

\bibitem{boyd2011distributed}
{S. Boyd, N. Parikh, E. Chu, B. Peleato, J. Eckstein, et~al.},
\newblock ``Distributed optimization and statistical learning via the
  alternating direction method of multipliers,''
\newblock {\em Foundations and Trends in Machine learning, vol. 3, no. 1, pp.
  1--122}, 2011.

\bibitem{biADMM2015}
G.\ Zhang and R.\ Heusdens,
\newblock ``Bi-alternating direction method of multipliers over graphs,''
\newblock in {\em Proc. Int. Conf. Acoustics, Speech, Signal Proc. pp.
  3571--3575}, 2015.

\bibitem{zhang2018distributed}
{G. Zhang and R. Heusdens},
\newblock ``Distributed optimization using the primal-dual method of
  multipliers,''
\newblock {\em IEEE Trans. Signal Process., vol. 4, no. 1, pp. 173--187}, 2018.

\bibitem{sherson2018derivation}
{T. Sherson, R. Heusdens, W. B. Kleijn},
\newblock ``Derivation and analysis of the primal-dual method of multipliers
  based on monotone operator theory,''
\newblock {\em IEEE Trans. Signal Inf. Process. Netw., vol. 5, no. 2, pp
  334-347}, 2018.

\bibitem{ryu2016primer}
{E. Ryu, S. P. Boyd},
\newblock ``Primer on monotone operator methods,''
\newblock {\em Appl. Comput. Math., vol. 15, no. 1, pp. 3-43,}, 2016.

\bibitem{giaconi2018privacy}
{G. Giaconi, D. Gündüz, H. V. Poor},
\newblock ``Privacy-aware smart metering: Progress and challenges,''
\newblock {\em IEEE Signal Process. Mag., vol. 35, no. 6, pp. 59-78}, 2018.

\bibitem{bogdanov2008sharemind}
{D. Bogdanov, S. Laur, J. Willemson},
\newblock ``Sharemind: A framework for fast privacy-preserving computations,''
\newblock in {\em Proc. 13th Eur. Symp. Res. Comput. Security: Comput.
  Security, pp. 192-206,}, 2008.

\bibitem{li2019privacyS}
{Q. Li and M. G. Christensen},
\newblock ``A privacy-preserving asynchronous averaging algorithm based on
  shamir's secret sharing,''
\newblock in {\em Proc. Eur. Signal Process. Conf., pp. 1-5}, 2019.

\bibitem{cover2012elements}
{T. M. Cover and J. A. Tomas},
\newblock {\em Elements of information theory},
\newblock John Wiley \& Sons, 2012.

\bibitem{Jane2020TIFS}
{Q. Li, J. S. Gundersen, R. Heusdens and M. G. Christensen},
\newblock ``Privacy-preserving distributed processing: Metrics, bounds, and
  algorithms,''
\newblock in {\em IEEE Trans. Inf. Forensics Security. vol. 16, pp.
  2090--2103}, 2021.

\bibitem{lopuhaa2019information}
{ M. Lopuha{\"a}-Zwakenberg, B. {\v{S}}kori{\'c} and N. Li},
\newblock ``Information-theoretic metrics for local differential privacy
  protocols,''
\newblock {\em arXiv preprint arXiv:1910.07826}, 2019.

\bibitem{Jane2020GSP}
{Q. Li, M. Coutino, G. Leus and M. G. Christensen},
\newblock ``Privacy-preserving distributed graph filtering,''
\newblock in {\em Proc. Eur. Signal Process. Conf., pp. 2155-2159}, 2021.

\bibitem{yagli2020information}
{S. Yagli, A. Dytso and H.V. Poor},
\newblock ``Information-theoretic bounds on the generalization error and
  privacy leakage in federated learning,''
\newblock in {\em IEEE 21st International Workshop on Signal Processing
  Advances in Wireless Communications (SPAWC)}, 2020, pp. 1--5.

\bibitem{bu2020tightening}
{Y. Bu, S. Zou, and V. V. Veeravalli},
\newblock ``Tightening mutual information-based bounds on generalization
  error,''
\newblock {\em IEEE J. Sel. Areas Info. Theory. vol. 1, no. 1, pp. 121--130},
  2020.

\bibitem{pensia2018generalization}
{A. Pensia, V. Jog, and P. L. Loh},
\newblock ``Generalization error bounds for noisy, iterative algorithms,''
\newblock in {\em Proc. of IEEE Int. Symp. Inform. Theory (ISIT), pp.
  546–550}, 2018.

\bibitem{negrea2019information}
Jeffrey Negrea, Mahdi Haghifam, Gintare~Karolina Dziugaite, Ashish Khisti, and
  Daniel~M Roy,
\newblock ``Information-theoretic generalization bounds for sgld via
  data-dependent estimates.,''
\newblock in {\em Proc. of Neural Information Processing Systems (NeurIPS), pp.
  11013–11023}, 2019.

\bibitem{cuff2016differential}
{P. Cuff and L. Yu},
\newblock ``Differential privacy as a mutual information constraint,''
\newblock in {\em Proc. 23rd ACM SIGSAC Conf. Comput. Commun. Secur., pp
  43--54}, 2016.

\bibitem{gotz2011publishing}
{ M. Gtz, A. Machanavajjhala, G. Wang, X. Xiao, J. Gehrke},
\newblock ``Publishing search logs—a comparative study of privacy
  guarantees,''
\newblock {\em IEEE Trans. Knowl. Data. Eng. vol. 24, pp. 520 - 532}, 2011.

\bibitem{haeberlen2011differential}
{A. Haeberlen, B. C. Pierce, A. Narayan},
\newblock ``Differential privacy under fire.,''
\newblock in {\em Proc. 20th USENIX Conf. Security., vol. 33}, 2011.

\bibitem{korolova2009releasing}
{A. Korolova, K. Kenthapadi, N. Mishra, A. Ntoulas},
\newblock ``Releasing search queries and clicks privately,''
\newblock in {\em Proc. Int'l Conf. World Wide Web, pp. 171--180}, 2009.

\bibitem{kifer2011no}
{D. Kifer and A. Machanavajjhala},
\newblock ``No free lunch in data privacy,''
\newblock in {\em SIGMOD, pp. 193--204}, 2011.

\bibitem{gaard2007Quan}
{J. Østergaard},
\newblock ``Multiple-description lattice vector quantization,''
\newblock in {\em Ph.D. dissertation, Delft University of Technology}. IEEE,
  2007.

\bibitem{schellekens2017quantisation}
{D. H. M. Schellekens, T. Sherson, and R. Heusdens},
\newblock ``Quantisation effects in {PDMM}: A first study for synchronous
  distributed averaging,''
\newblock in {\em Proc. Int. Conf. Acoust., Speech, Signal Process., pp.
  4237–4241}, 2017.

\bibitem{jonkman2018quantisation}
{ J. A. G. Jonkman, T. Sherson, and R. Heusdens},
\newblock ``Quantisation effects in distributed optimisation,''
\newblock in {\em Proc. Int. Conf. Acoust., Speech, Signal Process., pp.
  3649–3653}, 2018.

\bibitem{dolev1993perfectly}
{D. Dolev, C. Dwork, O. Waarts, M. Yung},
\newblock ``Perfectly secure message transmission,''
\newblock {\em J. Assoc. Comput. Mach., vol. 40, no. 1, pp. 17-47,}, 1993.

\bibitem{dall2002random}
{J. Dall and M. Christensen},
\newblock ``Random geometric graphs,''
\newblock {\em Physical review E, vol. 66, no. 1, pp. 016121}, 2002.

\bibitem{gupta2017privacy}
{N. Gupta, J. Katz, N. Chopra},
\newblock ``Privacy in distributed average consensus,''
\newblock {\em IFAC-PapersOnLine, vol. 50, no. 1, pp. 9515-9520}, 2017.

\bibitem{gupta2019statistical}
{N. Gupta, J. Kat and N. Chopra},
\newblock ``Statistical privacy in distributed average consensus on bounded
  real inputs,''
\newblock in {\em ACC, pp 1836-1841}, 2019.

\bibitem{li2019privacyA}
{Q. Li, I. Cascudo, and M. G. Christensen},
\newblock ``Privacy-preserving distributed average consensus based on additive
  secret sharing,''
\newblock in {\em Proc. Eur. Signal Process. Conf., pp. 1-5}, 2019.

\bibitem{nozari2017differentially}
{E. Nozari, P. Tallapragada, and J. Cort{\'e}s},
\newblock ``Differentially private average consensus: Obstructions, trade-offs,
  and optimal algorithm design,''
\newblock {\em Automatica, vol. 81, pp. 221--231}, 2017.

\bibitem{he2019privacy}
{J. He, L. Cai, C. Zhao, P. Cheng, X. Guan},
\newblock ``Privacy-preserving average consensus: privacy analysis and
  algorithm design,''
\newblock {\em IEEE Trans. Signal Inf. Process. Netw., vol. 5, no. 1, pp.
  127--138}, 2019.

\bibitem{ruan2017secure}
{M. H. Ruan, M. Ahmad, Y. Q. Wang},
\newblock ``Secure and privacy-preserving average consensus,''
\newblock in {\em Proc. Workshop Cyber-Phys. Syst. Secur. Privacy, pp.
  123-129}, 2017.

\bibitem{mo2017privacy}
{Y. Mo and R. M. Murray},
\newblock ``Privacy preserving average consensus,''
\newblock {\em IEEE Trans. Automat Contr., vol. 62, no. 2, pp. 753--765}, 2017.

\end{thebibliography}
\end{document}